\documentclass[aps,prd,reprint,superscriptaddress]{revtex4-2}

\usepackage[utf8]{inputenc}
\usepackage{amsmath}
\usepackage{amsfonts}
\usepackage{amssymb}
\usepackage{hyperref}
\usepackage{xcolor}

\begin{document}

\title{Gauging the scale invariance of Einstein equations: Weyl invariant \\ equations for gravity}

\author{Máximo Bañados}
\affiliation{Facultad de Física, Pontificia Universidad Católica de Chile,\\
Avenida Vicuña Mackenna 4860, Santiago, Chile}

\begin{abstract}
We derive a Weyl invariant equation for Gravity by gauging the global Weyl invariance of vacuum Einstein equations. The equation is linear in the curvature and a natural generalization of Einstein equations to Weyl geometry. The system has 5 physical polarizations, two tensor modes, two vectors modes and one scalar, associated to the cosmological constant. An exact black hole solution is found. We study the dynamics on Friedman backgrounds and the evolution of cosmological perturbations. It is shown that cosmological vector modes do not decay in this model.

\end{abstract}

\maketitle

\section{Introduction}
\label{intro}

H. Weyl introduced the concept of Weyl symmetry only two years after Einstein published his General theory of Relativity (GR) \cite{Weyl:1918ib}. He argued that a complete upgrade from flat space to curved space should allow rotations {\it and} dilatations for vectors parallel transported along a closed loop. The resulting structure, called Weyl geometry, incorporates a vector field $A_\mu$ via,
\begin{equation}\label{1}
\nabla_\mu g_{\alpha\beta} = 2 A_\mu\,  g_{\alpha\beta}.
\end{equation}
In Weyl geometry the metric $g_{\mu\nu}$ and the vector field $A_\mu$ are the geometrical data. The  most important property of  (\ref{1}) is invariance under Weyl transformations
\begin{eqnarray}
g'_{\mu\nu}(x) &=& e^{2\alpha(x)} \, g_{\mu\nu}(x),  \nonumber \\
A'_\mu(x) &=& A_\mu(x) + \partial_\mu \alpha(x) \nonumber\\
\Gamma'^{\mu}_{\ \nu\alpha} &=& \Gamma^{\mu}_{\ \nu\alpha}.  \label{tWeyl}
\end{eqnarray} 

With the information available back in 1918, Weyl interpret $A_\mu$ as the electromagnetic field. An immediate negative reaction came from Einstein himself who did not like the non-integrability of lengths of the new theory. This objection put a halt to an otherwise very interesting mathematical structure. Weyl idea has nevertheless received considerable attention. 

There exists several ways to achieve Weyl invariance at the level of the gravitational action. The most common frameworks are extensions/modifications of one of the following three models,
\begin{eqnarray}
I_1[g,A] &=& \int dx^4  \, \sqrt{g}\,g^{\alpha\mu}g^{\beta\nu} \hat R_{\alpha\mu} \hat R_{\beta\nu}, \nonumber\\
I_2[g] &=& \int dx^4 \, \sqrt{g}\, C^{\mu\nu\alpha\beta} C_{\mu\nu\alpha\beta}, \nonumber\\
I_3[g,\phi] &=&  \int dx^4  \, \sqrt{\tilde g}\, \tilde g^{\mu\nu}R_{\mu\nu}(\tilde g) \nonumber \\
&=&  \int dx^4  \, \sqrt{g}\, ( \phi^2 R + \partial_\mu\phi \partial^\mu\phi). \label{fake}
\end{eqnarray} 
The first one is Weyl's \cite{Weyl:1918ib} original proposal involving the square of the Ricci tensor built with the Weyl invariant connection satisfying (\ref{1}). This action depends on the metric and Weyl abelian field $A_\mu$. A second Weyl invariant action, also proposed by Weyl, is the $C^2$ action, called conformal gravity. This action depends only on the metric. The equations motion have higher derivatives. Third, the conformal trick (passing from Einstein frame to Jordan). If $\tilde g_{\mu\nu} = \phi^2 g_{\mu\nu}$, the metric $\tilde g_{\mu\nu}$ is trivially  invariant under appropriated transformation of $g_{\mu\nu}$ and $\phi$. It then follows that $I_3$ is Weyl invariant, simply because $\tilde g$ is invariant. (A quartic interaction $\phi^4$ can be added in this case). After more than 100 years of Weyl invariance the literature on this subject is massive and impossible to review here. See \cite{EPS,Romero_2012,Romero:2022sgv,Scholz:2012ev,Lobo_2018,Hobson_2022,Quiros:2022mut,Barcelo:2017tes,Berezin:2022odj,Haghani:2023nrm} for various discussions including Einstein's original objection, and \cite{scholz2017unexpected} for a historical review. Weyl symmetry without Weyl geometry has also received huge attention in the past. We mention the classic paper by Deser \cite{Deser:1970hs} and for more recent works \cite{Bars_2014,Ferreira:2016wem,shimon2022locally,Edery}, and references therein. See \cite{Jackiw:2014koa} for a criticism of the physical relevance of the conformal trick.

~

In this work we shall discuss a different Weyl invariant gravitational theory. The starting point is the invariance of the Einstein tensor,
\begin{equation}\label{G}
G_{\mu\nu}(g)=R_{\mu\nu}(g) - {1 \over 2} g_{\mu\nu}g^{\alpha\beta} R_{\alpha\beta}(g), 
\end{equation}
under {\it global} Weyl transformations, 
\begin{equation}\label{globalw}
g_{\mu\nu}(x) \rightarrow e^{2\alpha_0} \, g_{\mu\nu}(x).
\end{equation}
This follows directly from the invariance of $ g_{\mu\nu}g^{\alpha\beta}$ and the structure $\Gamma \sim {1 \over 2}g^{-1}\partial g $ of the Levi-Civita symbol. This symmetry implies that for any given solution of $G_{\mu\nu}(g)=0$, say $g_{(0)\mu\nu}$, there exists a one-parameter family of new solutions $g_{(\alpha_0)\mu\nu} = e^{2\alpha_0}  g_{(0)\mu\nu} $,  $\alpha_0 \in \Re$. For example, a mass $M$ Schwarzschild metric multiplied by $e^{2\alpha_0}$ is equivalent to a mass $e^{\alpha_0} M$ Schwarzschild metric (up to a trivial coordinate rescalling). See \cite{Garfinkle:1996kt}, for example, for a discussion on the role of this symmetry in relation to Choptuik scaling \cite{Choptuik:1992jv}.  

The symmetry (\ref{sl}) is not a symmetry of the Einstein-Hilbert action, but only of the equations of motion (\ref{ee0}). This situation is not new, the harmonic oscillator equation is invariant under $x(t)\rightarrow \sigma \, x(t) $ while the action is not. (In that example the space of solutions can be parametrized using the symmetry.) 

Our approach will be to consider the gauged version of (\ref{globalw}). We shall add a vector field $A_\mu$ and following the usual gauging procedure to find a locally Weyl invariant generalization of (\ref{G}). Now, before going to the local version of the scale symmetry we discuss the addition of a cosmological constant, that will have an important role in the formalism. 

In Sec. \ref{lambda} we add a cosmological term and explore how the whole procedure can be formulated in a language similar to spontaneous symmetry breaking. The analogy is purely mathematical, but it does illuminate some steps. In Sec. \ref{gauging} we turn on the vector field and write down the Weyl invariant equations. Again the language of spontaneous symmetry breaking is useful to understand the relevant degrees of freedom. In particular, we will see that the gauge field acquires a mass. In Sec. \ref{flat} we explore the perturbations around flat space showing that the theory has 5 polarizations. In. Sec. \ref{matter} we discuss one possible framework to add matter consistently with Weyl invariance.  In Sec. \ref{sphsec} we discuss solutions with spherical symmetry, and finally in Sec. \ref{cosmo} the basics of cosmology (Friedman and perturbations equations).

\section{Adding a cosmological constant}  
\label{lambda}
A cosmological constant added to (\ref{G}) would clearly hard-break the symmetry (\ref{globalw}). There is a simple workaround, however, with unexpected implications. Consider the equation,  
\begin{equation}\label{ee0}
G_{\mu\nu}(g) + \lambda(x)\, g_{\mu\nu} =0,
\end{equation}
where $\lambda(x)$ is a {\it field}. This equation has interesting features. First, (\ref{ee0}) determines $\lambda(x)$ since the contracted Bianchi identity implies $\partial_\mu\lambda=0$. The cosmological constant appears as an integration constant (see \cite{Feng:2024rnh} for a recent analysis of various approaches of this mechanism) and in this sense via spontaneous symmetry breaking. We shall elaborate this in detail below. 

A different, equivalent, interpretation of (\ref{ee0}) considers taking the trace and eliminating $\lambda(x)$ in terms of the scalar curvature. The result is the famous trace-free unimodular Einstein equation, 
\begin{equation}\label{unimodular}
R_{\mu\nu} - {1 \over 4} R g_{\mu\nu}=0. 
\end{equation}
Clearly (\ref{ee0}) implies (\ref{unimodular}). The converse it also true: (\ref{unimodular}) implies (\ref{ee0}), where, again, the cosmological constant appears as an integration constant. This procedure can be done with or without matter. See \cite{Weinberg:1988cp,Ellis:2010uc,Carballo-Rubio:2022ofy} for discussions on unimodular gravity and \cite{Henneaux:1989zc} for a constrained Hamiltonian analysis. Eq. (\ref{unimodular}) was also discussed in \cite{Hitchin:1982vry}, under the name `Einstein-Weyl' equation, in analogy with Einstein spaces. Some unexpected and interesting results have been discussed in \cite{KPTod_1996,Calderbank-1999,Bonneau:1999gj,Bonneau:1998wz,Bonneau:1998hd,Bonneau:1996jy,tod2021conformal} for other works. Recently, an action principle for (\ref{unimodular}), at $d=3$, was reported \cite{Klemm_2020}.  To our knowledge, the interpretation of (\ref{unimodular}) as a full Weyl invariant theory of gravity has not been discussed elsewhere.

A third interesting aspect of (\ref{ee0}) is that it preserves the  global Weyl symmetry. Since $\lambda(x)$ is not a fixed background field, Eq. (\ref{ee0}) is invariant under the simultaneous rescaling,  
\begin{eqnarray}
g_{\mu\nu}(x) &\rightarrow & e^{2\alpha_0} \, g_{\mu\nu}(x), \nonumber \\
\lambda(x) &\rightarrow & e^{-2\alpha_0} \lambda(x).  \label{sl} 
\end{eqnarray}
Of course this symmetry is broken spontaneously once a solution
\begin{eqnarray}\label{bg}
\lambda = \Lambda_0,
\end{eqnarray}
is chosen. The particular value $\Lambda_0=0$ would preserve the symmetry. This leads to two phases, the ``broken" phase $\Lambda_0 \neq0$, and the ``unbroken" phase $\Lambda_0=0$. 

A final useful property of (\ref{ee0}), the presence of the field $\lambda$ allows the definition of a Weyl invariant metric $\lambda g_{\mu\nu}$. This concept, developed in detail below, will be very useful throughout this work. 

~

The global Weyl symmetry just described, and its breaking, exhibits some similarities (and differences) with the Abelian $U(1)$ global Higgs model for a complex scalar $\Phi(x)$. Preparing the arena for the gauge version, it is worth elaborating this analogy.  

The $U(1)$  Higgs field can conveniently be expressed as $\Phi(x) = \rho(x) e^{i\theta(x)} $ with the global $U(1)$ symmetry acting additively $\theta(x)\rightarrow \theta(x) + \theta_0$ and leaving $\rho(x)$ invariant. Under symmetry breaking $\theta=\theta_0$ and $\rho(x)$ acquires a vev. 

Let us introduce analogous variables for our system. We replace the metric and scalar field $\{g_{\mu\nu}, \lambda\}$ by new fields $\{q_{\mu\nu}, \gamma\}$ related by 
\begin{eqnarray}
\lambda(x) &=& \Lambda_0 \,e^{-2\gamma(x)},  \\
g_{\mu\nu}(x) &=& q_{\mu\nu}(x) e^{2\gamma(x)}. \label{frd}
\end{eqnarray} 
In the new variables the global Weyl symmetry acts additively $\gamma \rightarrow \gamma+\alpha_0$ and $q_{\mu\nu}$ is invariant. The field $\gamma$ then plays the role of Goldstone boson and $q_{\mu\nu}$ will acquire a vev. $\Lambda_0$ is a constant needed for dimensional reasons. 

Using the well known expressions for the curvature of conformally related metrics, Eq. (\ref{ee0}) written in terms of $q_{\mu\nu}(x)$ and $\gamma(x)$ is, 
\begin{equation}\label{ee1}
\underbrace{G_{\mu\nu}(q) + \Lambda_0\, q_{\mu\nu}}_{\mbox{``vev equation"}} + \underbrace{H_{\mu\nu}(q,\gamma)}_{\mbox{``Goldstone"}} =0,
\end{equation}
with, 
\begin{equation}
H_{\mu\nu} = 2 _\mu\gamma \tilde D_\nu\gamma-2\tilde D_\mu \tilde D_\nu\gamma + q_{\mu\nu}( 2\tilde D_\alpha\tilde D^{\alpha}\gamma +\tilde  D^{\alpha} \gamma\tilde D^{q\alpha}\gamma). 
\end{equation}
Here $\tilde D_\mu$ is the $q_{\mu\nu}$-compatible covariant derivative $\tilde D _\mu q_{\alpha\beta}=0.$ The separation ``vev" plus ``Goldstone" equations can be explained as follows.

The contracted Bianchi identity $\tilde D^{\mu}G_{\mu\nu}(q)=0 $ acting on (\ref{ee1}) implies $\tilde D ^\mu H_{\mu\nu}=0$. After a short calculation it can be proved that this equation implies (as expected)
\begin{equation}\label{eqtheta}
\Lambda_0\,\partial_\nu\gamma(x)=0,
\end{equation}
in full consistency with $\partial_\mu\lambda(x)=0$, valid in the original variables.   

Once a value $\gamma=\gamma_0$, satisfying (\ref{eqtheta}), is chosen the symmetry  is broken. Next, for constant  $\gamma_{0}$ the combination $H_{\mu\nu}$ vanishes and (\ref{ee1}) becomes Einstein equations with a cosmological constant $\Lambda_0$. This is an equation for the ``vev" $q_{\mu\nu}$, independent of the Goldstone boson $\gamma(x)$. 

Despite this similarities, the Abelian Higgs model and global Weyl invariance of (\ref{ee0}) has nevertheless important differences. First of all the value of the vev is an integration constant and not a parameter in the action.   Second, for the $U(1)$ Abelian Higgs field, the complex phase $e^{i\theta} $ does not mix with radius $\rho$. Different values of $\theta$ represent rotations at fixed $\rho$. In our case, where the relevant group is $\Re$, a shift in $\gamma$ could be absorbed by a rescaling of $q_{\mu\nu}$.  We identify $\Lambda_0$ --introduced in (\ref{frd}) for dimensional reasons-- as the value of $\lambda$ at $\gamma=0$.

Theories with global Weyl invariance, and a non-trivial Weyl current, has recently been studied in \cite{Ferreira:2016wem,Ferreira:2016vsc,Ferreira:2018itt,Ferreira:2018qss}.

\section{Gauging the global Weyl invariance of Einstein equations}
\label{gauging}

In this section we study the gauged version of equation (\ref{ee0}) which brings in Weyl's geometry (\ref{1}) in a natural way. The analogy with the $U(1)$ Higgs model will see other implications, in particular, the Weyl vector $A_\mu$ will acquires a mass by ``eating" the Goldstone scalar mode. See \cite{Iorio:1996ad} for other aspects of gauging global Weyl transformations.

We start generalizing (\ref{tWeyl}) for arbitrary objects. Let ${\cal O}$ be a field transforming under Weyl as
\begin{equation}
{\cal O}' =  e^{p\alpha}  {\cal O}.
\end{equation}
We say that $p$ is the Weyl weight of ${\cal O}$.  The parameter $\alpha$ can be made local by turning on a vector field $A_\mu$ and replacing derivatives via the well-known trick
\begin{equation}\label{gaugingeq}
\partial_\mu {\cal O} \rightarrow \partial_\mu {\cal O} - p A_\mu {\cal O}.
\end{equation}
The gauge derivative satisfies, 
\begin{equation}
\partial_\mu {\cal O}' - p A'_\mu {\cal O}' = e^{p\alpha}  (\partial_\mu {\cal O} - p A_\mu {\cal O}),
\end{equation}
as needed to build quantities transforming like tensors under local Weyl transformations. Note that there is no imaginary unit ``$i$" here. The group of Weyl transformations is identified with $\Re$, not $U(1)$. 

Let us apply this procedure to Einstein equations. To avoid second derivatives, it is convenient to write Einstein's equations in first order form, 
\begin{eqnarray}
R_{\mu\nu}(\Gamma) - {1 \over 2} g_{\mu\nu}g^{\alpha\beta} R_{\alpha\beta}(\Gamma) + \lambda\, g_{\mu\nu}&=& 0,  \label{e01} \\   
\nabla_\mu g_{\alpha\beta}&=&0.  \label{e02}
\end{eqnarray}
This system is fully equivalent to (\ref{ee0}), in particular, is invariant under global Weyl transformations with the weights, 
\begin{eqnarray}
p(g_{\mu\nu}) &=& 2, \\
p(\Gamma^{\mu}_{ \ \nu\alpha} ) &=&0, \\
p(\lambda) &=& -2. 
\end{eqnarray} 
All derivatives inside the curvatures act on $\Gamma$, with $p=0$ thus they are no affected by the gauging process. The derivative appearing in  (\ref{e02}), on the other hand, does receive corrections by the gauging process. The ``gauged" version of equations (\ref{e01}) and (\ref{e02}) is then,
\begin{eqnarray}
R_{\mu\nu}(\Gamma) - {1 \over 2} g_{\mu\nu}g^{\alpha\beta} R_{\alpha\beta}(\Gamma) + \lambda\, g_{\mu\nu}&=& 0, \label{e1}   \\   
\nabla_\mu g_{\alpha\beta} - 2 A_\mu \, g_{\mu\nu}&=& 0,  \label{e2}
\end{eqnarray}
leaving the first equation intact. As a by-product we see that Weyl geometry, see Eq. (\ref{1}), can be understood as the gauging of the global Weyl symmetry of Einstein equations.

It is convenient to go back to second order formalism and eliminate the connection as an independent variable. This can be done because (\ref{e2}) is still algebraic for the connection. The solution to (\ref{e2}) is known as the Weyl connection \cite{Weyl:1918ib}
\begin{equation}
\hat \Gamma^{\mu}_{\ \alpha\beta}(g,A) = \Gamma^{\mu}_{\ \alpha\beta}(g) + A^\mu g_{\alpha\beta} - \delta^\mu_\alpha A_\beta  - \delta^\mu_\beta A_\alpha   \label{Wc}
\end{equation}
where $\Gamma(g)$ is the usual Levi-Civita form. From now on we denote by $\hat \Gamma(g,A)$ the solution to (\ref{e2}). In the same way its covariant derivative will be
\begin{equation}
\hat \nabla = d + \hat \Gamma
\end{equation}
The curvature is $\hat R^{\mu}_{\ \nu} = d\hat \Gamma^{\mu}_{\ \nu}+ \hat \Gamma^{\mu}_{\ \sigma}\wedge \hat \Gamma^{\sigma}_{\ \nu}$ and satisfies the Bianchi identity
\begin{equation}\label{bianchi0}
\hat \nabla \wedge \hat R^{\mu}_{\ \nu}=0.
\end{equation}

Having solved (\ref{e2}), Eq. (\ref{e1}) can be written in second order form 
\begin{equation}
\hat G_{\mu\nu}(g,A) +  \lambda\, g_{\mu\nu} =0, \label{main1}
\end{equation}
where $\hat G_{\mu\nu}(g,A)$ has the expansion (using (\ref{Wc})), 
\begin{eqnarray}
\hat G_{\mu\nu}(g,A) &=& G_{\mu\nu}(g) + D_\mu A_\nu +D_\nu A_\mu - 2 g_{\mu\nu} D_\alpha A^\alpha  \nonumber \\ 
&& \ \ \ \ \ \   +\ 2 A_\mu A_\nu +  g_{\mu\nu} A_\alpha A^\alpha. \label{hatG}
\end{eqnarray}  
Here, $D_\mu$ denotes the metric-compatible covariant derivative ($D_\mu g_{\alpha\beta}=0$) and $G_{\mu\nu}(g)$ is the metric Einstein tensor satisfying $D^\mu G_{\mu\nu}=0$.

Equation (\ref{main1}) will be our first (vacuum) master equation. In the following section we shall study several aspects of it.  Its main property is invariance under Weyl transformations, 
\begin{eqnarray}
g'_{\mu\nu}(x) &=& e^{2\alpha(x)} \, g_{\mu\nu}(x),  \nonumber \\
A'_\mu(x) &=& A_\mu(x) + \partial_\mu \alpha(x) \nonumber\\
\lambda'(x) &=& e^{-2\alpha(x)} \lambda(x) \label{tWeylf}
\end{eqnarray} 
Eq. (\ref{main1}) is second order in derivatives and connected continuously to Einstein equations. If $A_\mu$ is small enough, the extended Einstein tensor $\hat G_{\mu\nu}(g,A)$ is indistinguishable from $G_{\mu\nu}(g)$. 

Equation (\ref{main1}) could be written as, 
\begin{eqnarray}\label{ng}
G_{\mu\nu}(g) + \lambda g_{\mu\nu} &=& -\Big( D_\mu A_\nu +D_\nu A_\mu - 2 g_{\mu\nu} D_\alpha A^\alpha  \nonumber \\ && \ \ \ \ \ \   + 2 A_\mu A_\nu +  g_{\mu\nu} A_\alpha A^\alpha\Big) 
\end{eqnarray}
where we have passed all $A_\mu$-dependent terms to the right hand side. This is not a good idea because invariance of of $\hat G_{\mu\nu}(g,A)$ under (\ref{tWeyl}) occurs thanks to cancellations between the transformation of $g_{\mu\nu}$ and $A_\mu$. As can be seen, the right hand side of (\ref{ng}) is not gauge invariant, but the whole equation is invariant, thanks to terms coming from the left hand side. Equation (\ref{ng}) highlights the differences of the coupling between $g_{\mu\nu}$ and $A_\mu$ considered in this paper as compared to GR. We note, for example, that the $A_\mu$ ``energy-momentum" tensor has pieces linear in $A_\mu$.  Despite the differences, there will be some familiar relations too. As we now show, the field $A_\mu$ still satisfies Maxwell equations. 

Our problem now is to understand the dynamical content of (\ref{main1}). This equation contains three unknown fields, $g_{\mu\nu}, A_\mu$ and $\lambda$. Nicely, (\ref{main1}) is powerful enough to determine all of them. Just as $\partial_\mu\lambda(x)=0$ follows from (\ref{ee0}), we will see that (\ref{main1}) delivers Maxwell equations for $A_\mu$ and a scalar equation for $\lambda$. 

The calculation is simple. Start from (\ref{main1}) and apply a covariant derivative $D^{\mu}$. Using that $D^{\mu}G_{\mu\nu}=0 $ the  higher derivatives in  the metric disappear. Multiplying by $\sqrt{g}$ and after some standard manipulations one finds
\begin{equation}
\partial_\mu \Big(\sqrt{g} F^{\mu\nu}   \Big) +\sqrt{g} g^{\nu\sigma}\Big( \partial_\sigma \lambda + 2 A_\sigma \lambda\Big)   = 0,  \label{Maxwell0}
\end{equation}
that is, Maxwell equations with a source. Note that this equation is Weyl invariant 

Next, applying $\partial_\nu$ to (\ref{Maxwell0}) one finds the `Weyl invariant Laplacian' equation for $\lambda$,
\begin{equation}
{1 \over\sqrt{g} }\partial_\nu\Big( \sqrt{g}g^{\mu\nu}(\partial_\mu\lambda + 2 A_\mu \lambda) \Big) =0.     \label{scalareq}
\end{equation}

Eq. (\ref{Maxwell0}) describes the dynamics of a vector field and propagates 2 degrees of freedom. Eq. (\ref{scalareq}) describes a scalar and propagates one degree of freedom.  Summarizing, Eq. (\ref{main1}) propagates 5 physical degree of freedom: 2 tensors in $g_{\mu\nu}$, 2 vectors in $A_{\mu}$ and 1 scalar $\lambda$. This will be confirmed by a linear analysis of perturbations in Sec. \ref{flat}. It is interesting that $\lambda$ appears in (\ref{main1}) with no derivatives, yet it carries one physical degree of freedom.  It is worth stressing that (\ref{Maxwell0}) and (\ref{scalareq}) are not independent equations but consequences of (\ref{main1}).   In other words, the counting $5=2+2+1$ can be done directly from (\ref{main1}) without ever looking at (\ref{Maxwell0}) or (\ref{scalareq}). We shall do this in Sec. \ref{flat}. 

~ 

Going back to the analogy with the Abelian Higgs model we now show that the the Goldstone boson $\lambda$ can be eliminated ``get eaten" by the vector field $A_\mu$, giving it a mass. The five degrees of freedom are then arranged as 2 tensor polarizations plus a Proca field carrying three polarizations. In the unbroken phase with $\Lambda_0=0$, the Proca field becomes massless and there will be 4 degrees of freedom. 

Just like in the gauged Abelian Higgs model, the scalar mode $\lambda$ can be eliminated from the equations via a field redefinition. Consider the transformation $\{g_{\mu\nu},A_\mu,\lambda\}  \rightarrow \{q_{\mu\nu}, B_\mu, \gamma\} $ where,
\begin{eqnarray}
g_{\mu\nu} &= & q_{\mu\nu}\,  e^{2\gamma} ,\nonumber \\
A_\mu &= & B_\mu + \partial_\mu \gamma, \nonumber\\
\lambda &=& \Lambda_0\, e^{-2\gamma}. \label{fd} 
\end{eqnarray} 
In this transformation the field $\lambda(x)$ is replaced by  the field $\gamma(x)$. But this field redefinition can also be seen as a Weyl transformation with parameter $\gamma(x)$. Since $\hat G_{\mu\nu}$ is Weyl invariant, the result of acting on (\ref{main1}) with (\ref{fd}) is easy to compute,  
\begin{equation}
\hat G_{\mu\nu}(q, B) +  \Lambda_0\,q_{\mu\nu} =0.  \label{main2}
\end{equation}
We see that the field $\lambda$ (or $\gamma(x)$) has disappeared -got ``eaten" by the gauge field- leaving behind the scale $\Lambda_0$ acting as a cosmological constant.  Equation (\ref{main2}) is fully equivalent to (\ref{main1}).  They represent the same system. 

An alternative, simpler, way to arrive at (\ref{main2}) is via gauge fixing.  Equation (\ref{main1}) is Weyl invariant. This symmetry can be fixed by $\lambda = \Lambda_0$ transforming (\ref{main1}) into (\ref{main2}) (with $q_{\mu\nu}=g_{\mu\nu}$ and $B_{\mu}=A_{\mu}$).  

Repeating the cascade procedure that gave rise to Maxwell equations (\ref{Maxwell0}), we apply $D^{\mu} $ to (\ref{main2}) obtaining now the Proca equation for the field $ B$,  
\begin{equation}
\partial_\mu \Big(\sqrt{g} F^{\mu\nu}   \Big) +2\Lambda_0 \, \sqrt{g}\, B^\nu   = 0 \label{Proca1}
\end{equation}
where $F_{\mu\nu} = \partial_\mu B_\nu-\partial_\nu B_\mu = \partial_\mu A_\nu-\partial_\nu A_\mu $. In other words, $\lambda$ has disappeared and the gauge field has acquired a mass proportional $\Lambda_0$. 

The  parameter of a Weyl transformation depends arbitrarily on spacetime points. In this sense Weyl transformations define a `gauge symmetry' and their direct consequences are not observable. Observables must be invariant under the transformation (like electric and magnetic fields in electrodynamics). The metric $q_{\mu\nu}$ is invariant and can be postulated as the observable metric. We shall assume in this paper that observable quantities, like particles, couple to $q_{\mu\nu}$. This will be the main tool to couple matter to this theory. See Sec. \ref{matter}. 

Geodesic motion, for example, can be analysed in this way. Particles will not couple to $g_{\mu\nu}$ but to $q_{\mu\nu}$. If true, then the correct geodesic equation is built with the Levi-Cevita connection $\hat\Gamma^{\mu}_{\ \nu\alpha}(q)$ for the metric $q_{\mu\nu}$. There is a competition here because the Weyl connection $\hat\Gamma^{\mu}_{\ \nu\alpha}(g,A)$ in (\ref{Wc}) is also Weyl invariant. We have then two invariant connections and two geodesic equations. The Weyl connection would assume a coupling between particles and the Weyl vector $A_\mu$, which does not look very natural. This discussion is certainly an interesting point to elaborate, but it is beyond the scope of this work.

\section{Flat spectrum}
\label{flat}
The statement that $\lambda$ appearing in (\ref{main1}) carries one degree of freedom may seem counter-intuitive, since it appears with no derivatives in that equation. In this section we have a quick look at the linear spectrum of (\ref{main1}) expanding around the flat background $g_{\mu\nu}=\eta_{\mu\nu},\ A_\mu=0, \ \lambda=0$. This analysis will confirm the assertion of 5 degrees of freedom claimed before. There is also a nice Hamiltonian structure that is worth mentioning. 

Consider, 
\begin{eqnarray}\label{hs}
g_{\mu\nu} = \eta_{\mu\nu} + \epsilon\, h_{\mu\nu},  \ \ \ \ A_\mu =0 + \epsilon\, a_\mu, \ \ \ \  \lambda = 0+ \epsilon\,\delta, \nonumber
\end{eqnarray}
where $h_{\mu\nu}(x), a_{\mu}(x)$ and $\delta(x)$ are regarded as first order perturbations. We shall use a parametrization for the perturbations that will be useful when analysing Cosmology. Let 
\begin{eqnarray}
h_{\mu\nu}\!  &=& \!  e^{ikz}  \left( \begin{array}{cccc}
-2\psi & s_1 & s_2 & ik B \\ 
s_1 & -2 \phi + h_+ & h_\times & ik f_1 \\ 
s_2 & h_\times & -2 \phi - h_+ & ik f_2 \\ 
ik B & ik f_1 & ik f_2 &  - 2 \phi+2k^2 E
\end{array}  \right) \nonumber \\
a_\mu&=& e^{ikz}\, (a_0,a_1,a_2,ik \alpha),  \nonumber \\
\delta &=&  e^{ikz} \ell  \label{h}
\end{eqnarray} 
where it is assumed a wave with well-defined momentum $\vec{k}=k \, \hat z$. Here, all components of $h_{\mu\nu},\, a_\mu$ and $\ell$ are functions of time only. The  perturbations fall in the standard decoupled classes,

~

\centerline{
\begin{tabular}{l|c|c|c}
\ \ \ \ & $g_{\mu\nu}$ & $A_\mu$ & \ \ $\lambda$ \ \   \\ \hline
7  scalars&\ ${\psi,\phi,B,E,}$\ & $\ {a_0,\alpha}\ $ & $\ell$  \\ 
6 vectors modes& ${s_i,f_i,}$  &  $a_i$  & -  \\ 
2 tensor modes & ${h_\times},h_+$ & -   & -  \\ 
\end{tabular} }

~

Due to diff and Weyl invariance we can set 5 functions to convenient values. We will choose the Newton gauge $E=B=f_i=0$ supplemented with $\alpha=0$ to fix the Weyl gauge freedom. [Alternatively, one can express the equations in terms of observables (Bardeen variables) and $E,B,f_i,\alpha$ disappear. Either way the result is the same.] 

Eq. (\ref{main1}) imply the following dynamics. 

The 2 tensor modes follow the usual GR dynamics,
\begin{equation}
\ddot h_\alpha + k^2 h_\alpha =0, \ \ \ \ \ \ (\alpha=+,\times), \label{gw}
\end{equation}
totally expected since the new ingredient in this theory is a vector. The vector subspace does receive corrections. The equations are, 
\begin{eqnarray}
\dot s_i   &=& 2a_i ,   \\
\dot a_i &=&-{1 \over 2}k^2 s_i    \label{v1}
\end{eqnarray} 
Nicely, these equations can be collected in a Hamiltonian action, with a positive definite Hamiltonian,
\begin{equation}\label{Iv}
I_v[a_i,s_i] = \int dt \left( s_i\, \dot a_i - \Big( a_i^2 +  {1 \over 4} k^2 s_i^2 \Big) \right)
 \end{equation}
with $a_i, s_i$ as canonically conjugate pairs. The vector sector then contributes with 2 degrees of freedom. 

The scalar equations of motion are, 
\begin{eqnarray}
\ell + 2 k^2 \phi &=&0, \nonumber\\
a_0 + 2 \dot\phi &=&0, \nonumber\\
 k^2 (\phi-\psi) &=& 0, \nonumber\\
\ell + 2 \dot a_{0} + 2 \ddot\phi &=& 0. \label{scalarl}  
\end{eqnarray}  
Note that $\psi=\phi$ and both $\ell$ and $a_0$ can be expressed algebraically in terms of $\phi$ obtaining, 
\begin{equation}\label{scalarmode}
\ddot\phi + k^2 \phi=0.
\end{equation}
Thus, the scalar sector does carry one physical mode, as stated before. 

Note also that if we had set $\lambda=0$ in (\ref{main1}), then its fluctuation would vanish, $\ell=0$, and the scalar sector is totally trivial $\phi=\psi=a_0=0$. The scalar mode (\ref{scalarmode}) exists thanks to $\lambda(x)$ in (\ref{main1}). 

The scalar equations can also be collected as Euler-Lagrange equations for the functional
\begin{eqnarray}
I_s[\phi,\psi,a_0,\ell] = \int dt \left( -{1 \over 2}\dot\phi^2 + \phi \dot a_0 +{k^2 \over 4} \psi^2 +{k^2 \over 4}\phi^2 \right. \nonumber\\ \left. - {k^2 \over 2}\phi\psi + {1 \over 2}\ell\, \phi + {1 \over 8k^2}\ell^2 -{1 \over 4} a_0^2  \right). \ \ \    \label{Is}
\end{eqnarray}
and $\psi,a_0,\ell$ satisfy algebraic equations they can be integrated out leaving the physical mode
\begin{eqnarray}
I_{s,red}[\phi] = \int dt \left( {1 \over 2}\dot\phi^2 - {k^2 \over 2} \phi^2 \right).
\label{Isr}
\end{eqnarray}

Summarizing, the flat space perturbation analysis confirms the existence of 5=2+2+1 physical degrees of freedom: 2 tensor, 2 vector and 1 scalar. We emphasize that these statements follow solely from Eq. (\ref{main1}), which contains all the dynamical information. The same conclusions would arise from (\ref{main2}). This time the splitting is 5=2+3 (2 tensor, 3 massive vector).   

To the best of our knowledge $I_v$ and $I_s$ do not descend in any obvious way from a full non-linear covariant action. We have only shown that the linearized equations can be collected as variational equations for $I_v, I_s$. Note also that (\ref{Is}) contains a non-local part. This goes away, however, when integrating out $\ell$.

\section{Matter}
\label{matter}

We now include matter fields in the analysis. As we have stressed, we are not using the action as the guiding principle to build interactions, but only the equations of motion. Equations (\ref{main1}) or (\ref{main2}) do not descend from any obvious action principle, yet they are quite reasonable and have interesting physics. Our goal now is to add an energy momentum tensor to our equations, keeping Weyl invariance. 

An example could be a massless free scalar field whose energy momentum tensor, 
\begin{equation}
T^{\mbox{(scalar)}}_{\mu\nu}=\partial_\mu\phi\partial_\nu\phi - {1 \over 2} g_{\mu\nu} g^{\alpha\beta} \partial_\alpha\phi \partial_\beta\phi.
\end{equation}
is Weyl invariant.  Maxwell's energy momentum tensor, on the other hand, is not Weyl invariant despite the action being invariant. 

Rather than looking for specific cases we shall explore a generic picture allowed by the Weyl invariant fields $B_\mu,q_{\mu\nu}$ defined in (\ref{fd}) and that satisfy (\ref{main2}). Basically, we shall assume that all observables --like particles-- couple to the Weyl invariant metric $q_{\mu\nu}$. In this way we avoid several conceptual problems that arise when talking about Weyl invariant interactions.  Then, instead of working with (\ref{main1}), we shall now consider (\ref{main2}) as the starting point. 

Let us generalize (\ref{main2}) to 
\begin{equation}
\hat G_{\mu\nu}(g,A) +  \Lambda_0\, g_{\mu\nu} = 8\pi\, T_{\mu\nu}   \label{main}
\end{equation}
where we recall, $\hat G_{\mu\nu}(g,A)$ is given in (\ref{hatG}). The reason to have the energy-momentum tensor at the right hand side is to be as close as possible to GR. Indeed, setting $A_\mu$ to zero this equation is identical to Einstein equations. Of course, our interest is to explore the effects of $A_\mu$.    

The first challenge is to determine the correct conservation law for $T_{\mu\nu}$. Recall that $\hat G_{\mu\nu}$ is not conserved.  Let us  apply $D^{\mu} $ to (\ref{main}). After a short calculation one finds, 
\begin{equation}
\partial_\mu \Big(\sqrt{g} F^{\mu\nu}   \Big) +2 \sqrt{g}\, \Lambda_0 A^\nu   = 8 \pi\,\sqrt{g} J^\nu, \label{Proca2}
\end{equation}
where we have introduced the abbreviation, 
\begin{equation}
J_\nu=\hat\nabla^\mu T_{\mu\nu}.    \label{DT} 
\end{equation}
and $\hat \nabla$ is the Weyl covariant derivative (\ref{Wc}). 

Imposing $\hat\nabla^\mu T_{\mu\nu}=0$ is not necessarily well motivated. $\hat\nabla_\mu$ is not the Levi-Cevita covariant derivative and Eq. (\ref{main}) does not imply such condition. To keep the analysis general, we shall leave $J_\mu$ as a new ingredient. We will need more input to determine this current.  From (\ref{Proca2}), one natural requirement is 
\begin{equation}
\partial_\nu(\sqrt{g} J^\nu )=0, \label{DJ}
\end{equation} 
This is one equation, we will need more conditions to fully determine $J^\mu$ (see below). 

Summarizing so far, in this theory, $T_{\mu\nu}$ is the source for $g_{\mu\nu}$, while $J_\mu$ is the source for $A_\mu$. The two sources are related by (\ref{DT}).

We now explore some consequences of this system.

\section{Spherical symmmetry}
\label{sphsec}

We start with black holes solutions with $T_{\mu\nu}=0$. These can be studied analytically. The field,  
\begin{eqnarray}
ds^2 &=& - f(r)dt^2 +  {dr^2 \over f(r)(1 + k^2 r^2)}  + r^2\,d\Omega^2 \nonumber \\
A &=& k\,  f(r) dt       \label{sph}
\end{eqnarray} 
with
\begin{eqnarray}
f(r) &=& 1 - {2M  \over r}\sqrt{1 + k^2 r^2} \label{sch}  \\ 
&& +{W \over kr} \left(-kr +  \sqrt{1 + k^2 r^2 } \ln\left(k r  +\sqrt{1 + k^2 r^2} \right) \right) \nonumber   
\end{eqnarray}
with
\begin{equation}
W = -{\Lambda_0 \over k^2} 
\end{equation}
is an exact solution to (\ref{main}). This solution has two integration constants $M$ and $k$. The constant $k$ has dimensions 1/length and sets an extra scale. A salient feature is that the metric (\ref{sch}) approaches the Schwarzschild metric for {\it small} $r$. For large $r$ the potential grows logarithmically $f(r) \sim 1 + W \log(kr)$.

The metric (\ref{sph}) has black hole features. For negative $\Lambda_0$ and positive $M$ a horizon $f(r_+)=0$ always exists. Horizon regularity can be checked by going, for example, to Eddington-Finkelstein coordinates 
\begin{equation}
dt = du \pm {\sqrt{h} \over f} dr
\end{equation}
rendering both $g_{\mu\nu}$ and $A_\mu$ regular. 

For $M=0$, the metric (\ref{sph}) is regular at $r=0$. The ``ground state" $M=W=0$ has constant Riemann spatial curvature (it is also conformally flat) and the gauge field is constant $A = k dt$.  

If (\ref{sph}) is interpreted as the exterior geometry of a regular object (localized $T_{\mu\nu}$), one would like to relate $M$ to it's energy-momentum tensor. We will not attack here the full issue of conserved charges that poses several challenges. We will restrict the discussion to static, spherically symmetric configurations. We follow \cite{Abbott:1981ff} closely, with modifications needed to our case.  

The starting point is the Weyl-geometry contracted Bianchi identity \cite{WOS:000313064700013,Straub-2015}
\begin{equation}\label{BW}
\hat\nabla^\mu (\hat G_{\mu\nu} - F_{\mu\nu}) =0. 
\end{equation}
This identity is proved following exactly the same steps as the usual one: starting from (\ref{bianchi0}) and contracting twice with the metric tensor (paying attention to the non-metricity of $\hat\nabla$). 

Multiplying (\ref{BW}) by an arbitrary vector field $P^\nu$ (and $\sqrt{g}$) and integrating by parts the identity (\ref{BW}) is transformed into, 
\begin{equation}
\partial_\mu( \sqrt{g}(\hat G^{\mu\nu} - F^{\mu\nu})  P_\nu) =   \sqrt{g}(\hat G^{\mu\nu} - F^{\mu\nu})\hat  \nabla_\mu P_\nu.  \label{cons1}
\end{equation}
We emphasize that so far the vector $P^\mu$ is arbitrary.  If the right hand side was zero, then we would have a conserved current. 
 
We now split (\ref{sch}) into a background field, defined by $M=W=0$, plus fluctuations (terms proportional to $M,W$). Observe that $\Lambda_0$ is treated as first order. It is easy to see that  $\hat G_{\mu\nu}-F_{\mu\nu}$ is of first order then all other pieces in (\ref{cons1}) can be evaluated on the background. In particular, the covariant derivative at the right hand side of (\ref{cons1})  is evaluated on the background. 

Without assigning any geometrical interpretation to $P_\mu$ let us assume it satisfies, 
\begin{equation}
\hat \nabla_\mu P_\nu = k\, P_{\mu} P_\nu. \label{Pcond}
\end{equation}
(The factor $k$ is needed for dimensional reasons. The numerical coefficient is irrelevant.) Eq. (\ref{Pcond}) has two virtues. First, it eliminates $F^{\mu\nu}$ in the right hand side of (\ref{cons1}), a piece difficult to evaluate in the star interior. Second, it has a static, spherically symmetric solution, 
\begin{equation}
P_\mu dx^\mu = \, dt - {kr \over 1 + k^2 r^2 } dr.  \label{Psol}
\end{equation}
(One may wonder why not imposing $\hat \nabla_\mu P_\nu=0$. This would kill the right hand side of (\ref{cons1}) and leave a conserved current. The problem is that this condition does not have static, spherically symmetric solutions for $P^\mu$. We thus impose the weaker condition (\ref{Pcond}).)

Inserting (\ref{Psol}) into (\ref{cons1}), integrating over $d^3x$ (regulated at a large distance $L$), using (\ref{main}), and assuming static fields the following relation follows,
\begin{eqnarray} \label{M}
8\pi  \int d^3x \sqrt{g}\, T^{\mu\nu} P_\mu P_\nu &=&{1 \over k} \int_\infty d\Omega \sqrt{g} (G^{r\nu}-F^{r\nu})P_{\nu} +  \nonumber  \\ 
&& +\, \Lambda_0  \int d^3x \sqrt{g}\, P^\mu P_\mu .
\end{eqnarray}
The angular integral at $r\rightarrow\infty$ is evaluated at the known exterior solution (\ref{sch}). The second term is an integral over all space. Since $\Lambda_0 \sim W$ is first order, the integral can be evaluated at the background over all space. It turns out that both terms in the right hand side of (\ref{M}) diverge logarithmically (for large $r$), but the sum is convergent and equal to $-8\pi M$. We thus find the relation,
\begin{eqnarray}
\int d^3x \sqrt{g}\, T^{\mu\nu} P_\mu P_\nu = -M,
\end{eqnarray}
relating the integration constant $M$ with the matter energy-momentum, as desired. For $k=0$, the usual formula 
\begin{equation}
\int  d^3x \,\sqrt{g} \ T^{0}_{\ 0}=M
\end{equation}
is recovered.

\section{Cosmology}
\label{cosmo}

We now study (\ref{main}) in the context of cosmology.  Our goal is to exhibit the Friedman and perturbations equations associated to (\ref{main}), in particular, to explore the consequences of the current $J^\mu$ appearing in  (\ref{Proca2}). As it happens in GR, analytical solutions are available in some situations. In this paper we do not explore the phenomenological viability of  these models. 

Consider the `background + perturbations'  fields
\begin{eqnarray}
ds^2 &=& a(\eta)^2 \eta_{\mu\nu}dx^\mu dx^\nu  + \epsilon a(\eta)^2 \, h_{\mu\nu}(x)\ dx^\mu dx^\nu  \nonumber\\
A &=&  A_0(\eta)d\eta +  \epsilon\, a_\mu(x)\ dx^\mu  \nonumber\\
J &=& J_0(\eta) d\eta + \epsilon j_\mu(x)\ dx^\mu 
\end{eqnarray}
where the perturbations are the same as in flat space, see Eq.(\ref{h}), again, assuming a wave with $\vec{k}=k \hat z$. We only need to add the parametrization for $j_\mu$, 
\begin{equation}
j_{\mu} = e^{ik z} ( j_0(\eta), j_1(\eta),j_2(\eta), ik \varphi(\eta)),
\end{equation}
Matter is assumed as an ideal fluid 
\begin{equation}
T_{\mu\nu} = (p+\rho) u_\mu u_\nu + p g_{\mu\nu}, 
\end{equation}
where all quantities are splited into background + perturbations following standard notations.  

The independent background equations are
\begin{eqnarray}
{3 \over a^2}W^2 &=& 8\pi \bar \rho + \Lambda_0  \label{fr1} \\
\bar \rho'+ 2A_0 \bar \rho + 3W (\bar \rho + \bar p)  &=&  -J_0 \label{fr2} \\
\Lambda_0A_0 &=& 4\pi J_0, \label{w1} \\
a J_0'+ 2 a'J_0 &=& 0,   \label{w2}
\end{eqnarray} 
where $\bar \rho$ and $\bar p$ refer to the background values of energy density and pressure respectively. Eq. (\ref{w1},\ref{Proca2}) follow from (\ref{w2},\ref{DJ}), respectively, and we have introduced the abbreviation, 
\begin{equation}
W = {a' \over a} - A_0 \label{W}
\end{equation}
which plays the role of Hubble parameter everywhere, including the perturbation equations. 

The solution to Eqns. (\ref{w1}) and (\ref{w2}) is:
\begin{equation}
A_0 = {\kappa \over a^2}, \ \ \ \ \ \  \ \  J_0 = {\Lambda\,   \over 4\pi} {\kappa \over a^2},  \label{solc0}
\end{equation} 
where $\kappa$ is an integration constant that will characterize the departure from GR Friedman models. Nicely, analytic solutions exists for the background fields (for all $w$). We display here the matter and radiation examples (in the early universe when $a\ll 1$, and $\Lambda_0a^2$ in (\ref{fr1}) can be ignored),

~

\centerline{
\begin{tabular}{c|c|c}
& matter & radiation \\ 
\hline
$a(\eta)$ & $a_m \eta^2 \sqrt{1 - {2\kappa \over 3 a_m^2 \eta^3 }}$ &  $a_r \eta \sqrt{1 - {2 \kappa \over a_r^2\eta} }$ \\ 
$\bar \rho(\eta)$ & ${9 \over  6\pi a_m^2\, \eta^6 -4\pi\kappa\eta^3} $ & ${3  \over 8\pi a_r^2\, \eta^4 -16\pi\kappa \eta^3}$ \\
$W(\eta)$ &  ${2 \over \eta}$ &  ${1 \over \eta}$  
\end{tabular}} 

~

The constants $a_m/a_r$ measure the amount of matter/radiation, respectively. If $\kappa>0$ then the Big-Bang ($a\rightarrow 0 ,\bar \rho \rightarrow \infty$) is reached at $\eta=2\kappa/a_r^2$ with a different scaling. Passing to proper time, near $a=0$, we find $a\sim t^{1/3}$.   

We now discuss the evolution of perturbations. We use Newton's gauge $E=B=f_i=0$. 

\underline {Tensor modes:}  As expected (since the new ingredient is a vector),  the tensor perturbations satisfy the same equations as in GR, 
\begin{equation}
h''_{\alpha} + 2 W h'_{\alpha} + k^2 h_\alpha =0, \ \ \ \ \ \ \ \ \   (\alpha=+,\times)
\end{equation}
where $W$ is the background function (\ref{W}).

\underline{Vector modes:} The metric vector modes couple to $A_\mu$ and exhibit a different evolution as compared to GR. The independent equations governing the evolution of $s_i,a_i,\delta u_i$ are
\begin{eqnarray}
{1 \over 2}k^2 s_i - 2 W a_i + a_i' &=& 8 \pi a^2 (\bar \rho+\bar p) \delta u_i \label{cv1} \\
s_i'+ 2 W s_i   &= & 2 a_i \label{cv2} \\
a_i'' + (k^2 - 2 \Lambda_0a^2) a_i &=& - 8\pi a^2 j_i \label{cv3}
\end{eqnarray} 

Two comments are in order before solving this system. First, Eq. (\ref{cv3}) contains the vector part of the current $J_\mu$. We do not have a theory for the values of this current.  For the moment, as a first approximation, we shall simply set $j_i=0$ in (\ref{cv3}). Second, we shall omit the piece $\Lambda_0a^2$ in (\ref{cv3}) assuming an early universe --before acceleration. With these conditions the system can be solved trivially.

Eq. (\ref{cv1}) determines $\delta u_i$ in terms of the other variables. Eq. (\ref{cv3}) has a simple solution, 
\begin{equation}
a_i(\eta) = D_{i} \sin(k\eta)  \label{ai}
\end{equation}
with $D_i$ an integration constant (we have set a second constant to zero to keep expressions short). We now solve (\ref{cv2}) during matter domination,  
\begin{eqnarray}
s_i &=& {C_i \over \eta^4} + {D_i \over \eta^4}\Big[(24-12k^2\eta^2+k^4 \eta^4)\cos(k\eta) + \nonumber \\  
&& \ \ + 4 k\eta(-6+ k^2 \eta^2)\sin(k\eta) \Big]  
\end{eqnarray}
with $C_i$ a new integration constant. The first term, proportional to $C_i$, is exactly the GR evolution, decaying as $1/\eta^4$. The second part shows a drastic difference with GR. For late times, $s_i \sim D_i k^4\cos(k\eta)$ has a constant amplitude oscillation, off phase with respect to $a_i$. Note however, that this evolution is controlled by the independent parameter $D_i$.

\underline{Scalar modes:} Before displaying all equations we mention that $A_\mu$ induces anisotropic stress in the system. Indeed, the Einstein-Weyl equations imply,
\begin{equation}\label{alpha}
\psi - \phi = 2\alpha,
\end{equation} 
where $\alpha$ is the longitudinal part of $a_\mu$ (see (\ref{h})). On a first approximation we will set $\alpha=0$. This condition has an extra motivation. The perturbation $j_\mu$ contributes with two unknowns to the scalar sector ($j_0$ and $\varphi$), and so far we have produced only one equation for them, Eq. (\ref{DJ}). Setting (\ref{alpha}) gives the needed condition to have a closed system.  Of course this needs a better understanding, linked to the nature of the current $J_\mu$. We hope to come back to this in the future. 

Taking into account (\ref{alpha}), the remaining independent 6 equations for $\phi,\delta\rho,q,a_0,j_0,\varphi$ are: 
\begin{eqnarray}
-k^2 \phi -3 W (\phi' + W \phi)  &=&  4 \pi a^2 \delta\rho + 3 W\, a_0 \nonumber \\
 W \phi + \phi' &=& 4\pi a^2 q -  {1 \over 2}a_0   \nonumber \\
\phi'' + 3 W\, \phi' +a^2(\Lambda-8\pi \bar p) \phi &=& 4\pi a^2\delta p -  a_0' -W\, a_0   \nonumber \\
-(k^2-2\Lambda_0a^2) a_0 &=& 8 \pi a^2 j_0 \nonumber \\
a_0' &=& 8 \pi a^2 \varphi \nonumber  \\
\kappa \Lambda_0\phi'- 2\pi \kappa j_0 &=& \pi a^2 \left( j_0'  + 2W j_0  + k^2 \varphi \right) \nonumber
\end{eqnarray}
$q$ is related to the longitudinal part of $u_{\mu}$ \cite{Baumann_2022}. The first three equations follow directly from (\ref{main}). The other three follow from (\ref{Proca2}) and (\ref{DJ}). These are simpler than (\ref{DT}), but of course consistent with it. Note that $W$, defined in (\ref{W}), plays again the role of Hubble parameter.

Setting the Weyl dependent functions, $a_0,j_0,\varphi$ and $\kappa$, to zero this system reduces to the standard GR perturbation equations. We explore now the full system on the simple equation of state $p=w \rho$. 

The last three equations imply a simple differential equation for $a_0$ whose solution is
\begin{equation}
a_0(\eta)= {4 \kappa  \over a(\eta)^2}(\phi(\eta)-D_0)
\end{equation}
with $D_0$ an integration constant. Next, $\delta\rho, q,j_0$ and $\varphi$ can be solved algebraically as functions of $\phi$ and one is left with a master equation for the potential
\begin{equation}
 \phi''+P_1\, \phi'+ P_0 \, \phi + Q_0  =0
\end{equation}
with
\begin{eqnarray}
P_1 &=& 3(1+w)W + {4\kappa \over a^2 } \nonumber\\
P_0 &=& w k^2 - {8 \kappa^2\over a^4} + (1+w)\Lambda_0 a^2 + {4(-1+3w)\kappa W \over a^2} \nonumber\\
Q_0 &=& {4D_0 \kappa(2\kappa + (1-3w)a^2 W  \over a^4} 
\end{eqnarray}  
Exact solutions for this equation are available on some cases. For example on matter dominance ($w=0$ and $a^2\Lambda_0=0$) the solution is
\begin{equation}
\phi(\eta) = C_0 + {4\kappa (D_0-C_0) (3a_m^2 \eta^3+\kappa)  \over (3a_m^2 \eta^3 -2\kappa)^2 } + { C_1(6a_m^2 \eta^3 - \kappa) \over \eta^2(3a_m^2 \eta^3 -2\kappa)^2}, \nonumber
\end{equation}
where $C_0, C_1$ are new integration constants. At $\kappa =0$ the GR form $\phi = C_0 + \tilde C_1/\eta^5$ is recovered. 

For general $w$, a simple equation for the density contrast can be derived,
\begin{equation}
\left( {\delta \rho \over \bar \rho} \right)' = 3(1+w) \phi' + {4\kappa(1+3w)(\phi - D_0) \over a^2}.
\end{equation}   
Here the first term is the usual GR contribution.

\section{Conclusions}
 
Summarizing, we have built a Weyl invariant equation fro gravity by gauging the global invariance of the Einstein tensor.  We have explored the linear spectrum of (\ref{main1}) clarifying the number of propagating degrees of freedom. We have found and exact black hole and study Friedman models. Analytic expressions are presented, and shown to differ from GR expressions via adjustable integration constants. Two important departures from GR are (i) the spherically symmetric Newton potential grows logarithmically at large $r$, (ii) cosmological vector modes do not decay at late times.

The existence of a Lagrangian for this theory is an open question. Given the experience at $d=3$ \cite{Klemm_2020}, this task is probably not simple. The linearized actions (\ref{Is}) and (\ref{Iv}), with positive definite Hamiltonians, are interesting, but their relationship to a covariant, interacting action needs a detailed study.

\acknowledgments
 
The author would like to thank G. Barnich, R. Bravo, A. Faraggi, L. Garay, G. García, P. Ferreira, A. Gomberoff,  M. Henneaux, S. Rica and S. Theisen for many enlightening conversations. This work was partially funded by Fondecyt Grant (Chile) \# 1201145.


\end{document}